# Generation and Application of Bessel Beams in Electron Microscopy

Vincenzo Grillo[1,2], Jérémie Harris[3], Gian Carlo Gazzadi[1], Roberto Balboni[4], Erfan Mafakheri[5], Mark R. Dennis[6], Stefano Frabboni[1,5], Robert W. Boyd[3], Ebrahim Karimi[3]

1 CNR-Istituto Nanoscienze, Centro S3, Via G Campi 213/a, I-41125 Modena, Italy
2 CNR-IMEM Parco Area delle Scienze 37/A, I-43124 Parma, Italy
3 Department of Physics, University of Ottawa, 25 Templeton St., Ottawa, Ontario K1N 6N5, Canada
4 CNR-IMM Bologna, Via P. Gobetti 101, 40129 Bologna, Italy
5 Dipartimento di Fisica Informatica e Matematica, Università di Modena e Reggio Emilia, via G Campi 213/a, I-41125 Modena, Italy
6 H.H. Wills Physics Laboratory, University of Bristol, Bristol BS8 1TL, United Kingdom

**Abstract**

We report a systematic treatment of the holographic generation of electron Bessel beams, with a view to applications in electron microscopy. We describe in detail the theory underlying hologram patterning, as well as the actual electro-optical configuration used experimentally. We show that by optimizing our nanofabrication recipe, electron Bessel beams can be generated with efficiencies reaching 37±3%. We also demonstrate by tuning various hologram parameters that electron Bessel beams can be produced with many visible rings, making them ideal for interferometric applications, or in more highly localized forms with fewer rings, more suitable for imaging. We describe the settings required to tune beam localization in this way, and explore beam and hologram configurations that allow the convergences and topological charges of electron Bessel beams to be controlled. We also characterize the phase structure of the Bessel beams generated with our technique, using a simulation procedure that accounts for imperfections in the hologram manufacturing process. Finally, we discuss a specific potential application of electron Bessel beams in scanning transmission electron microscopy.

**Introduction**

Electron vortex beams have recently drawn significant attention within the electron microscopy community, and have shown great potential for a host of applications [1,2]. For example, electron vortex beams have recently been produced with orbital angular momenta as large as $200\hbar$ per electron; such beams show promise for potential applications in magnetic measurement [3]. For this reason, a great deal of effort has been expended in attempts to optimize the efficiency of vortex beam generation. In particular, holographic elements have emerged as promising candidates for high efficiency structured electron beam generation [4-9].

Holographic optical elements can allow electron beams to be shaped by modulating the transverse phase and amplitude profiles of incident electron waves with high precision. Amplitude modulation of incident electron beams can be achieved by alternating thick fringes made from opaque material with regions of high transparency. By contrast, phase modulation is carried out by varying the transverse thickness profile of a nearly transparent material, so as to produce disparities in the electro-optical path lengths experienced by different transverse components of the incident beam [5,6].

Phase-modulating elements have already found a range of applications in electron microscopy [10-12]. Specifically, phase plates can be used in transmission electron microscopy (TEM) to improve the contrast of weak phase objects, or to compensate for spherical aberration effects [13]. Attempts have also been made to produce phase plates for scanning transmission electron microscopy (STEM), in one case resulting in a Fresnel lens analogous to zone plate lenses for X-rays [14]. However, these types of lenses pose a significant nanofabrication challenge.

Beyond the examples mainly focused on vortex beams, relatively little work has been done with a view to shaping electron beams using holographic elements [5][8][9], and still less with reference to specific practical applications. This is not to suggest that this area is entirely unexplored; studies have previously investigated silicon nitride (SiN) as a candidate holographic material for electron beam shaping, for its low electron-optical density, and its ability to modify beam phase directly in axis [15]. However, no medium, no matter how transparent, can ever act as a perfect phase plate, since atoms in the material can always produce inelastic or high-angle scattering that can, in essence, be treated as absorption. This scattering can represent a significant hindrance to the use of on-axis phase holograms, producing a "frosted glass" effect, which results in a blurring of the transmitted beam, and a reduction in its quality [16]. The use of SiN holograms for on-axis electron beam shaping faces another drawback, in that it requires that thickness modulations be applied with precisions on the nanometer scale, a significant challenge even using state-of-the-art nanofabrication techniques.

In this sense, the introduction of off-axis amplitude holograms can be considered a significant development. These holograms, which consist of a modulated diffraction grating, benefit from the absence of unwanted scattering from their transparent regions by alternating fully absorbing and fully transparent fringes. A second advantage to this approach is that the phase imprinted on the incident wavefront is encoded in the transverse grating profile, and is therefore readily controlled, even when imperfect manufacturing techniques are employed. This technique does suffer from an important drawback, however, in that it typically results in low-efficiency generation of the desired output beam. Recently, we introduced off-axis phase holograms that allow this limitation to be overcome, potentially reaching efficiencies as large as 100% [5,6]. Here, we report a detailed study of electron Bessel beam generation using this technique.

Bessel beams are widely used in photonics, and have recently been discussed theoretically in the context of a number of electron microscopy applications. In the ideal case, Bessel beams are possess a propagation-invariant profile, and are therefore referred to as diffraction-free modes (see the discussion in Section 3). These beams hold great promise for their ability to reduce channeling [17], to control aberrations and potential applicability to new imaging modes, as well as for the generation of optical tractor beams, and other exotic applications. Apart from their wide range of potential applications, Bessel beams have also drawn considerable interest on theoretical grounds, for their unusual properties [18]. It has been noted that these beams could be applied to fundamental studies of beam polarization [19], since the efficiency of orbit-spin conversion for a Bessel beam could in principle reach 100%.

Notably, electron beams of approximately Bessel form have been generated using on-axis techniques such as hollow cone illumination [20]. However, electron beams generated in this way suffer from large intensity losses due to the partial blocking of the beam required by the technique. Still more critically, this strategy does not allow for the modification or control of key beam parameters, such as topological charge and convergence.

Here, we report a detailed study of the first off-axis Fresnel phase hologram to generate electron Bessel beams [5], and examine: 1) the conditions under which Bessel beams can be generated and applied to microscopy and imaging; 2) techniques by which key beam and hologram parameters, including topological charge, transverse wavenumber, and hologram aperture radius can be adjusted; and 3) the main practical limitations of electron Bessel beam generation.

## 1. Holographic Generation of Structured Electron Beams

Holographic plates can be used to confer spatial structure upon arbitrary electron beams with high efficiency. These devices are fabricated by inducing spatially varying changes in the optical thickness and transmissivity of a material, and therefore amount to optical phase and amplitude masks. When an incident plane wave is transmitted through such a mask, it gains a position-dependent phase $\Delta\varphi(\rho,\phi)$ relative to a reference wave having travelled an identical distance in vacuum, and experiences a spatial amplitude modulation $A(\rho,\phi)$, such that the mask may be described by a transmittance

$$T(\rho,\phi) = A(\rho,\phi)e^{i\Delta\varphi(\rho,\phi)} \quad (1),$$

where $\rho,\phi$ are the standard cylindrical coordinates. The transverse wavefunctions $\psi_{in}(\rho,\phi)$ and $\psi_t(\rho,\phi)$, respectively corresponding to the incident and transmitted beams, are then related by $\psi_t(\rho,\phi) = T(\rho,\phi)\psi_{in}(\rho,\phi)$. Three nontrivial classes of hologram may be distinguished, with reference to Equation (1). First, *phase holograms* are those for which $\Delta\varphi(\rho,\phi)$ exhibits a spatial dependence, while the hologram's amplitude modulation function is spatially constant, i.e. $A(\rho,\phi) = A_0$. By contrast, *amplitude holograms* induce a spatially varying amplitude modulation, but a spatially constant phase in the incident beam, so that $\Delta\varphi(\rho,\phi) = \Delta\varphi_0$. Finally, *mixed holograms* are characterized by spatially varying phase and amplitude modulations, so that neither $A(\rho,\phi)$ nor $\Delta\varphi(\rho,\phi)$ is spatially constant for these masks.

In what follows, we shall restrict our attention to phase holograms, which may in general be associated with a transmittance $T(\rho,\phi) = A_0 e^{i\Delta\varphi(\rho,\phi)}$. Physically, the phase modulation $\Delta\varphi(\rho,\phi)$ is induced in the incident beam due to the inner potential $V(\rho,\phi,z)$ of the material from which the holographic mask is constructed. This potential results in the addition of an energy term $eV(\rho,\phi,z)$ to the total Hamiltonian governing the time evolution of the electron beam in the material, resulting in a phase shift of the transmitted beam, relative to a reference wave having travelled the same distance in vacuum. From the general solution to the relativistic Schrödinger equation, this phase shift is found to be

$$\Delta\varphi(\rho,\phi) = C_E \int_0^{t(\rho,\phi)} V(\rho,\phi,z)\,dz, \quad (2)$$

where $t(\rho,\phi)$ is the variation in the thickness of the hologram as a function of position in the transverse plane, and $C_E = \frac{2\pi e}{\lambda}\frac{E+E_0}{E(E+2E_0)}$ is a constant for a particular electron kinetic energy $E$, rest energy $E_0$, and de Broglie wavelength $\lambda$. In our case, the inner potential of the phase mask may be approximated by its mean value, $V_0$, such that [21][22]

$$\Delta\varphi(\rho,\phi) = C_E V_0 \int_0^{t(\rho,\phi)} dz = C_E V_0 t(\rho,\phi). \quad (3)$$

Hence, an arbitrary transverse phase profile can be imprinted on the incident beam, provided that variations in the local phase mask thickness $t(\rho,\phi)$ can be controlled with sufficient precision.

## 2. Generation and Propagation of Bessel Beams

We shall now focus our attention specifically on the generation of electron Bessel beams, which are described by scalar wavefunctions of the form

$$\Psi(\rho,\phi,z;t) = J_n(k_\rho\rho)e^{in\phi}e^{-i(\omega t - k_z z)}, \quad (4)$$

where $J_n$ represents an $n^{th}$ order Bessel function of the first kind, $n$ is an integer, $k_\rho$ and $k_z$ are respectively the wavefunction's transverse and longitudinal wave vector components, and $\omega$, the electron's angular frequency, is related to its de Broglie wavelength $\lambda$ by $k^2 = k_\rho^2 + k_z^2 = \frac{2m\omega}{\hbar} = \left(\frac{2\pi}{\lambda}\right)^2$, where $k$ is the modulus of the electron wavevector and $\hbar$ is the reduced Planck constant. These beams carry an amount of orbital angular momentum (OAM) along their propagation direction given by $L_z = n\hbar$ per electron, as indicated by the presence of a phase term $e^{in\phi}$ in the expression (4).

The generation of a Bessel beam necessarily entails imprinting a phase of the form $\Delta\varphi = \beta = k_\rho\rho + n\phi$ onto the incident wavefunction. This can be achieved by choosing a phase hologram with transmittance $T(\rho, \phi) = A_0 e^{i\beta}$. An additional grating term $k_x x = k_x \rho \cos\phi$, where $k_x = \frac{2\pi}{\Lambda}$ and $\Lambda$ is a grating constant, can also be introduced to $\beta$ for later convenience, so that

$$\beta = k_\rho \rho + n\phi + k_x x. \quad (5)$$

A functionally identical hologram, for which the imprinted phase becomes $\Delta\varphi = \text{Mod}(\beta, 2\pi)$, where $\text{Mod}(a, b)$ represents the remainder obtained when dividing $a$ by $b$, would be equally well-suited to generating Bessel beams. We refer to this latter phase mask, in which $T(\rho, \phi) = A_0 e^{i\,\text{Mod}(\beta, 2\pi)}$, as a *blazed hologram*. Although blazed holograms are optimal from the standpoint of maximizing the efficiency of Bessel beam generation, they are difficult to produce in practice due to the finite resolution of existing fabrication techniques, which make use of a limited number of imprinted pixels to produce phase masks. As a result, the ideal blazed holograms must often be approximated by alternative configurations. In particular, by choosing the experimentally achievable phase imprint function $\Delta\varphi = \varphi_0 \cos(\beta)$, Bessel beams may be generated without prohibitively low efficiency. Phase masks of this form are referred to as *sinusoidal holograms*. From Equation (3), we note that in this case $\varphi_0 \cos(\beta) = C_E V_0 t(\beta)$, so that in practice, these holograms can be produced by inducing sinusoidal variations $t(\beta) = \frac{1}{2} t_0 \cos\beta$ in the mask thickness, where $t_0$ is the peak-to-valley thickness of the holographic material. Sinusoidal holograms are characterized by transmittance functions of the form

$$T(\rho, \phi) = A_0 e^{i\varphi_0 \cos(\beta)}. \quad (6)$$

Hence, the wavefunctions associated with the incident and transmitted electron beams are related by $\psi_t(\rho, \phi) = e^{i\varphi_0 \cos(\beta)} \psi_{in}(\rho, \phi)$. The Jacobi-Anger expansion may be applied to the exponential term to obtain $e^{i\varphi_0 \cos(\beta)} = \sum_{m=-\infty}^{\infty} i^m J_m(\varphi_0) e^{im\beta}$, where $m$ is an integer, so that upon substitution of Eq. (5),

$$\psi_t(\rho,\phi) = \psi_{in}(\rho,\phi) \sum_{m=-\infty}^{\infty} i^m J_m(\varphi_0) e^{im(k_\rho \rho + n\phi + k_x x)}. \quad (7)$$

For the case of a planar incident electron wavefunction of the form $\Psi_{in}(\rho,\phi,z;t) = e^{-i(\omega t - k_z z)}$, we have $\psi_{in}(\rho,\phi) = 1$, and therefore one obtains for the total transmitted wavefunction

$$\Psi_t(\rho,\phi,z;t) = e^{-i(\omega t - k_z z)} \sum_{m=-\infty}^{\infty} i^m J_m(\varphi_0) e^{im(k_\rho \rho + n\phi + k_x x)}. \quad (8)$$

Each term in the above expansion contains a component $e^{imk_x x} = e^{i\frac{2\pi m}{\Lambda} x}$, so that the transmitted wavefunction consists of an infinite number of diffracted beams, spaced apart at angles $\theta_m = m\frac{k_x}{k}$. We refer to $m$ as the order of diffraction, and note that the $m^{th}$-order diffracted beam will carry an OAM of $mn\hbar$, and will be characterized by a conical phase front $\propto e^{imk_\rho \rho}$. Further, the transmitted electron beam will be split among the various diffraction orders, with the $m^{th}$ order receiving a fraction $|J_m(\varphi_0)|^2$ of the total transmitted intensity. Maximally efficient generation of the beam associated with the $m^{th}$ diffraction order would therefore require that a value of $\varphi_0$ be chosen such that $|J_m(\varphi_0)|^2$ be maximized. Immediately after the hologram, the wavefunction associated with the $m^{th}$ diffracted beam takes the form $\Psi_t^m(\rho,\phi,z;t) = e^{i(m(k_\rho \rho + n\phi) + k_z z - \omega t)}$, where the z axis is now taken to lie along the propagation direction of the particular diffraction order in question.

While the electron wavefunction $\Psi_t^m(\rho,\phi,z;t)$ does not take the form of a Bessel function immediately after the holographic mask, it can be shown (see Appendix I and ref [23]) to take on Bessel character within a range of propagation distances given by $\rho \ll z \leq \frac{kR}{mk_\rho}$. In this region, we have for the transverse wavefunction

$$\psi_t^m(\rho,\phi) \approx N e^{i\left(kz - \frac{k\rho^2}{2z} + mn\phi + \frac{m^2 k_\rho^2 z}{2k}\right)} J_{mn}(mk_\rho \rho) \quad (9),$$

where $N$ is a dimensionless normalization constant.

We may additionally consider the far-field electron wavefunction, which describes the beam after the hologram in the region $z \to \infty$. It can be shown (see Appendix II) that under these conditions, the wavefunction assumes the form

$$\psi_t^m(k) \to e^{imn\phi} \delta(k - mk_\rho) \quad (10).$$

Theoretical Fresnel (near-field) and Fraunhofer (far-field) intensities associated with a Bessel beam generated from a phase hologram are displayed in Figure 1.

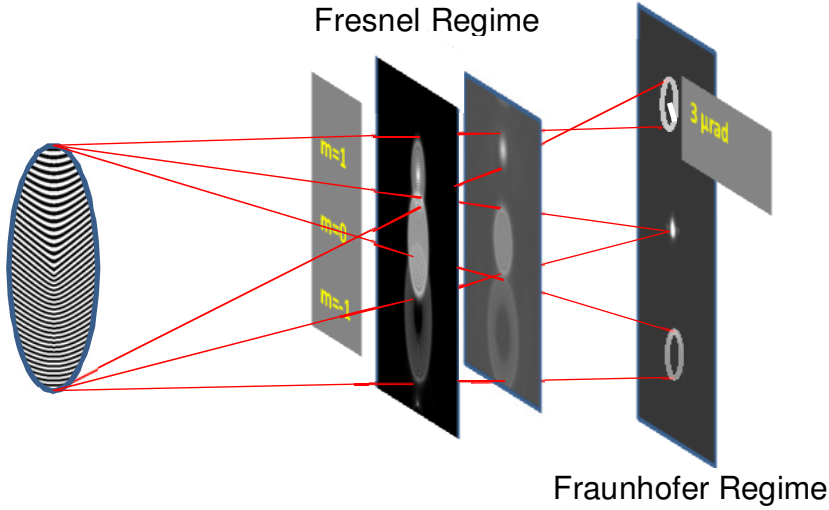

**Figure 1.** Theoretical Fresnel (near-field) and Fraunhofer (far-field) intensity distributions associated with a Bessel beam generated from an off-line phase hologram. The Bessel beam itself is formed at the first diffracted order ($m = 1$), and is found to take a ring-like form in the far-field, in accordance with Equation (10).

## 3. Properties of Bessel Beams

Bessel beams of the form (4) are solutions to the scalar wave equation, which in vacuum is given by

$$\left(\nabla^2 - \frac{1}{c^2}\frac{\partial^2}{\partial t^2}\right)\Psi(\rho, \phi, z; t) = 0. \quad (11)$$

This can readily be observed by substituting a trial solution in cylindrical coordinates of the form $\Psi(\rho, \phi, z; t) = R_n(\rho)e^{i(n\phi + k_z z - \omega t)}$ into Eq. (11), whence we find that

$$\rho^2 \frac{d^2 R_n(\rho)}{d\rho^2} + \rho \frac{dR_n(\rho)}{d\rho} + \rho^2 \left(\frac{\omega^2}{c^2} - k_z^2 - n^2\right) R_n(\rho) = 0,$$

which has solution $R_n(\rho) = J_n(k_\rho \rho)$, where $k_\rho^2 = \frac{\omega^2}{c^2} - k_z^2$ [24]. It then follows that $\Psi(\rho, \phi, z; t) = J_n(k_\rho \rho)e^{i(n\phi + k_z z - \omega t)}$, in agreement with (4). We note also that the transverse amplitudes of Bessel beams, $R_n(\rho)$, are independent of the beam propagation distance $z$. For this reason, Bessel beams are referred to as *non-diffracting* beams [25,26]. Despite their attractive physical properties, Bessel beams of the form (4) are not normalizable, carry infinite energy, and are therefore unphysical. Nonetheless, they can be closely approximated in practice, as we shall see. In Figure 2, we illustrate the non-diffractive propagation of an ideal Bessel beam, along with its propagation range, $z_{\max}$, which depends on the hologram convergence angle $\alpha \equiv k_\rho/k$.

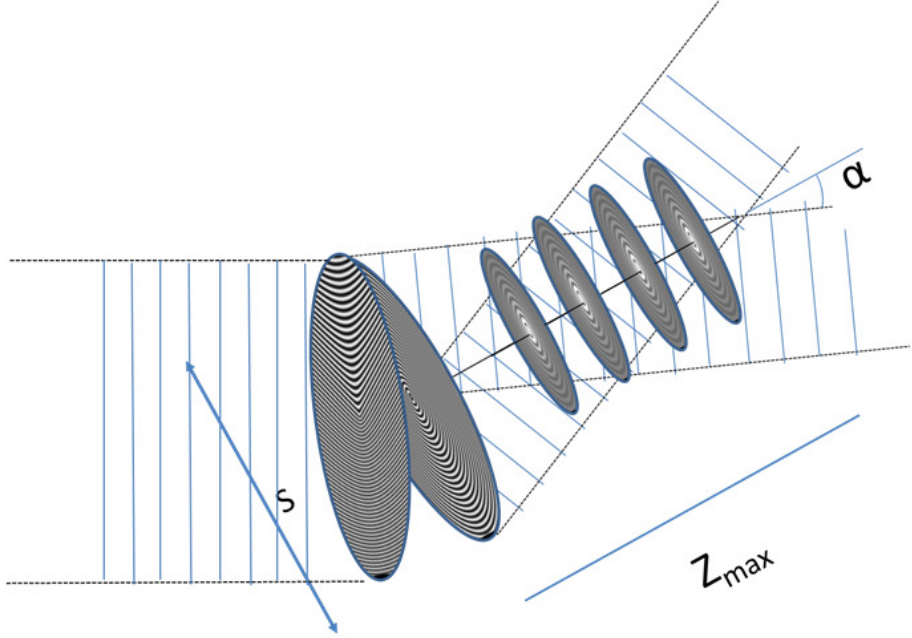

**Figure 2.** Theoretical depiction of diffraction-free propagation of an idealized electron Bessel probe. The diffraction-free propagation of the beam persists until it reaches the propagation range $z_{\max}$, beyond which the condition $z \leq \frac{kR}{mk_\rho}$ is no longer satisfied, and the beam loses its Bessel character. Notice also that the shape of the hologram projected in the diffracton direction is similar to that of a Fresnel lens.

## 4. Simulation of Electron Beam Propagation

Beam propagation simulations were carried out numerically using STEM_CELL software [27], which allows electron beam wavefunctions to be deduced based on our experimentally constructed hologram thickness maps. The electron wavefunction could then be calculated at different propagation distances by making use of the relation [28]

$$\psi(z + \Delta z) = P(\Delta z) \otimes \psi(z),$$

where $P(\Delta z)$ is the Fresnel propagator, describing the beam's evolution over a distance $\Delta z$, and $\psi(z)$ is the electron wavefunction at position $z$, which serves as a pupil function in the context of the Fresnel propagation integral, and $\otimes$ represents the convolution operation (see Appendix I). In the paraxial approximation, the propagator takes the form $P(\Delta z) = -\frac{i}{\lambda \Delta z} e^{\frac{i\pi\rho^2}{\lambda \Delta z}}$.

In practice, the electron wavefunction incident on the hologram is not perfectly collimated. For this reason, the aperture function $\psi(z)$ describing the incident beam is characterized by a slightly convergent wavefront. This requires that numerical simulations be carried out with a pixel size significantly smaller than the electron beam diameter.

We note that much of the blurring observed in the Bessel beams generated experimentally was due to the limited transverse coherence length of the source, brought upon by the finite size of the FEG Schottky emitter. This coherence length depends on the demagnification of the source at the sample plane. We accounted for limitations in transverse beam coherence by considering the Fresnel diffraction zone to be described by many mutually incoherent beams, each of which is characterized by a slightly different incidence angle [28].

From this work, it is therefore clear that the generation of truly propagation-invariant Bessel beams is limited in efficiency by three considerations. First, Bessel beams generated in the laboratory are characterized by intensity oscillations at beam center throughout propagation, due to diffraction from the hologram aperture. Second, the limited range of applicability of the approximation scheme used to derive the near-field electron wavefunction Equation (9) predicts the breakdown of Bessel-like behaviour at some maximal propagation distance, $z_{max}$. Indeed, well beyond this point, the wavefunction takes its far-field form Equation (10), and loses all Bessel character. Finally, imperfections in hologram patterning can result in non-ideal, pseudo-Bessel beams. Great care must therefore be taken to ensure that an optimal hologram design is chosen, so as to produce high-quality beams.

## 5. Hologram Patterning

Transmittance electron microscopy (TEM) experiments were primarily performed using a JEOL 2200FS microscope, equipped with Schottky field emission gun (FEG), operated at 200 keV. The hologram was inserted in the microscope's sample position, and beam images were obtained under low magnification, using the objective minilens as a Lorentz lens. This allowed for a large camera length and focal range, permitting imaging from the Fresnel to the Fraunhofer planes. This working mode, and the Fresnel mode in particular, are not calibrated in our microscope. As a result, we implemented a manual calibration scheme. The microscope was equipped with an Omega filter for energy loss imaging, and used to map hologram thickness profiles.

For scanning transmission electron microscopy (STEM) experiments, the hologram was mounted in the second condenser aperture of an FEI Tecnai TEM equipped with a Schottky FEG, and operated at 200 keV. A Dual-Beam instrument (FEI Strata DB235M), combining a focused gallium-ion beam (FIB) and a scanning electron microscope (SEM), was used to pattern the holograms by FIB milling 200 nm-thick silicon-nitride membranes coated with a 120 nm-thick gold film. The membranes were coated with the gold film in order to prevent electron transmission in all but the patterned areas.

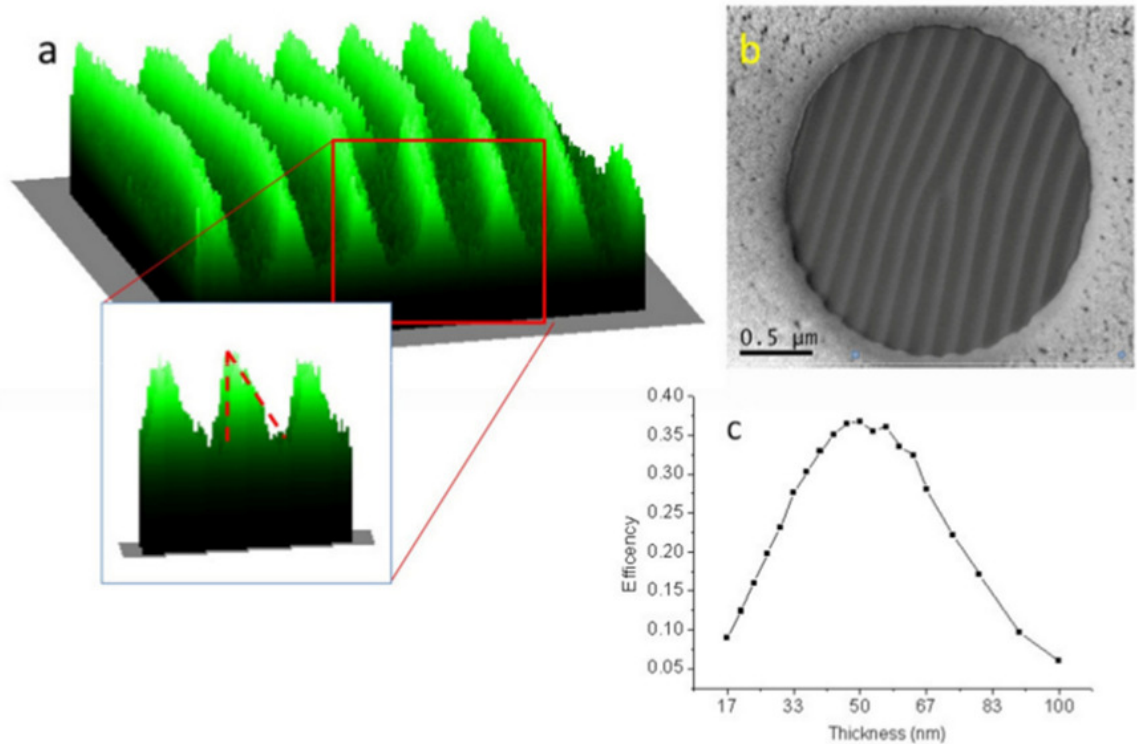

**Figure 3.** Experimental hologram patterning. **a** Three-dimensional rendering of an energy filtered TEM-based thickness map of the center of a hologram with parameters $n = 1$, $k_\rho = 3.2 \cdot 10^{-5}$ Å$^{-1}$, and $R = 1.22$ µm. **b** SEM image of the same hologram **c** Simulation of hologram efficiency as a function of thickness scaling factor $t_0$. For the thickness profile considered, a peak-to-valley thickness of 50nm is found to result in a maximal efficiency of 38%. We determined the efficiency of the hologram experimentally to be 37%.

We note that, at an operating potential of 200 keV, a 120 nm gold layer is not sufficient to completely stop the electron beam used. However, the presence of the gold film does suffice to induce elastic scattering of electrons. The mean free path associated with diffuse scattering is on the order of 60 nm while dynamic scattering occurs on the order of some tens of nanometers. Fortunately, the diffraction angles we explore are on the order of µrad (the Bragg angle for a grating with 100 nm step spacing), so that almost any scattering event produces a deviation from the angular range of interest. We found that, in practice, some detectable intensity was transmitted in the forward direction only when the beam was completely concentrated in one point. We experimentally determined the undesired forward transmittance to be well below 1%.

The procedure for hologram nanofabrication is implemented by starting with a bitmap picture of a computer-generated hologram, which is converted into a FIB pattern file containing three key pieces of information. These are respectively the pixel coordinates at which the FIB is switched on, the beam dwell time on each pixel, and the repetition number of the whole coordinate set, adjusted in such a way as to obtain the desired milling depth [6].

The second step is to adjust the FIB magnification according to the desired dimensions of the hologram. We selected a 50 nm width, and 100 nm periodicity for the stripes composing the hologram, resulting in a typical full hologram size on the order of 10 $\mu$m x 10 $\mu$m.

Once the computer generated hologram has been designed, the holograms are patterned, in two stages: first, the gold layer is uniformly removed from a circular region, 10 $\mu$m in diameter. To this end, the power transmitted from the secondary electron beam is monitored during milling, until a signal is observed, indicating that the gold is no longer present. Next, the hologram pattern is superimposed on the uncovered region, and milled into the silicon nitride.

For reasons related to the finite pixel resolution accommodated by our software, we imprinted the ideal, blazed profile only onto holograms with large grating periods, and nearly sinusoidal profiles onto those with grating periods under $\sim 300$ nm. In order to control the experimental hologram thickness profile, we performed TEM energy loss analyses. Through imaging, and by comparing beam transmission spectra, we generated quantitative maps of sample thickness.

The result of this procedure is shown in Figure 3, where we aimed to generate a sawtooth hologram profile. The inset shows that the thickness profile indeed corresponds closely to that of a blazed hologram. We can define the exit efficiency $\eta$ of the hologram as follows:

$$\eta = \frac{I_{m=1}}{\sum_m I_m},$$

where $I_m$ represents the intensity associated with the $m^{th}$ diffraction order. We note that this definition of efficiency differs from more canonical definitions, in that it explicitly considers beam intensities $I_m$ *after* transmission through the hologram, rather than providing the ratio of desired beam intensity to the intensity of the beam incident on the hologram aperture [8]. While these two definitions coincide in the limit of a strictly non-absorbing hologram, they will not agree in general, and from the known absorption of SiN were estimated to differ roughly by a factor of two to three in our experiment. This disagreement may be understood to arise from loss of beam intensity due to the absorption of electrons by the hologram.

Using this profile we can plot the hologram's efficiency as a function of the peak-to-valley thickness of the holographic material, $t_0$, from which we can see (Figure 3-c) that this profile allows a maximum efficiency of 38%. We obtained an efficiency of 37%, which is presently the best performance achieved by such a device, given that an uncertainty of about 3% must be allowed in order to account for the unknown intensity of the beams outside the field of view.

This also indicates that it is not possible to further increase the efficiency of this nanofabrication recipe; greater control of the groove profile is therefore necessary, but lies outside the scope of this work.

## 6. Results and Discussion

In presenting the data, we distinguish between two classes of hologram, based upon their respective aperture radii $R$. This parameter determines the extent to which the electron probe will resemble an ideal Bessel beam. Large aperture radii allow for the generation of highly Bessel-like beams in the Fresnel region, whereas reductions in $R$ lead to a decrease in the number of visible rings associated with the electron beam, all else being equal. It can also be shown that the aperture radius is inversely proportional to the width of the momentum distribution, such that $\Delta k 2R \approx 1$.
Thus, holograms with large apertures tend to produce ideal, delocalized Bessel beams suitable for interferometry, while smaller aperture holograms generate highly localized beams that are best suited to imaging.

In particular, we note that, for a small $k_\rho$ and large aperture $R$, the first-order diffracted beam will closely approximate a Bessel beam, whereas for smaller $R$ the hologram will predominantly act as a pinhole, resulting in significant overlap between the zeroth and first-order diffracted beams. We note also that, in the Fresnel regime, increases in aperture size do not increase the convergence of the generated beam.

### I. Bessel Beams with Large Aperture Radii

Figure 4-a,b shows two holograms, characterized by respective hologram convergence $k_\rho/k$ of 6 $\mu$rad and 15 $\mu$rad, and large, identical aperture sizes. Figure 4-c,d shows the corresponding Bessel-like beams generated from these holograms in the Fresnel region, when they are illuminated by approximately collimated incident electron beams. The holograms were prepared with $n = 0$, and therefore impart no OAM to the transmitted electron beams. Both holograms were 10 $\mu$m in diameter and contained 100 grating lines.

The Bessel beams shown in Figure 4-c,d reveal the critical role played by the radial wavenumber $k_\rho$ in defining the spread and number of visible fringes in the transmitted beams. For holograms with smaller values of $k_\rho$, the first-order diffracted beams are subject to relatively insignificant spreading during propagation, and the Bessel beams generated from these masks are therefore readily isolated from the zeroth diffracted order. By contrast, holograms manufactured with larger $k_\rho$ produce strongly divergent transmitted beams, resulting in significant overlap between the zeroth and first orders of diffraction, though this overlap can be reduced by increasing the main separation $k_x$. Indeed, the extent of this overlap can be so significant that the isolation of the first diffracted order from the zeroth order becomes challenging (Figure 4-d). This overlap also results in the apparent deformation of the first-order diffracted beam at its center. Holograms manufactured with small $k_\rho$ are also found to produce Bessel beams with fewer rings than would be the case for those manufactured with larger transverse wavenumbers, as expected theoretically. Hence, for a given aperture size, an increase in $k_\rho$ will result in a more Bessel-like electron beam in the Fresnel near-field, with a greater number of visible fringes.

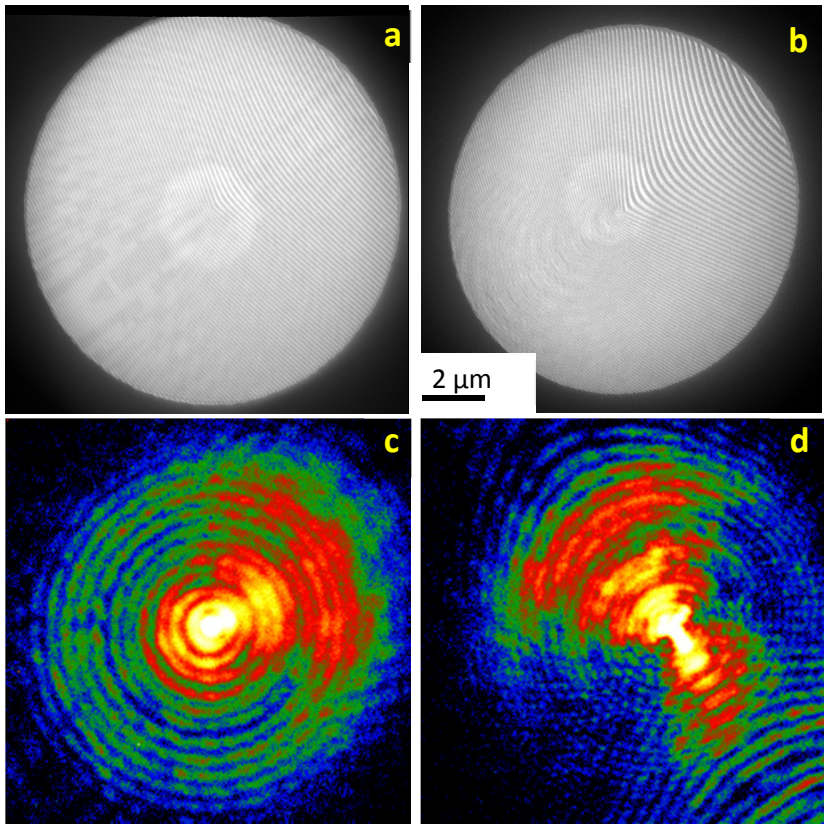

**Figure 4.** Bessel beam generation by large aperture phase holograms. **a** Scanning electron microscope (SEM) image of a phase hologram with aperture radius $R = 5\ \mu m$ and convergence angle $\alpha = 6\ \mu \text{rad}$. **b** SEM image of a phase hologram with aperture radius $R = 5\ \mu m$ and large convergence angle $\alpha = 15\ \mu \text{rad}$. **c** Near-field intensity pattern obtained experimentally from the hologram depicted in part **a**. **d** Near-field intensity pattern obtained from the hologram depicted in part **b**.

## II. Bessel Beams with Small Aperture Radii

For comparison, we show in Figure 5 a series of holograms manufactured with smaller aperture radii, along with corresponding intensity profiles for the first diffracted orders of the transmitted electron beam. In the figure, we compare the cases $n = 0, 1, 2$. In each case, the holograms were manufactured with a hologram convergence $\alpha = k_\rho/k$ of $1\,\mu\text{rad}$. Notably, in the case of $n = 2$, we reach an efficiency of almost $37 \pm 3\%$ which is by far the largest value ever achieved.

Under these conditions, the beam consists only of a very faint ring about the beam center, and its shape depends strongly on propagation distance. This can be understood to occur as a consequence of the small hologram aperture, which does not allow higher order fringes to manifest themselves upon propagation, resulting in a beam with almost no Bessel character. Such beams are well suited to imaging, owing to their small spot size and strong localization.

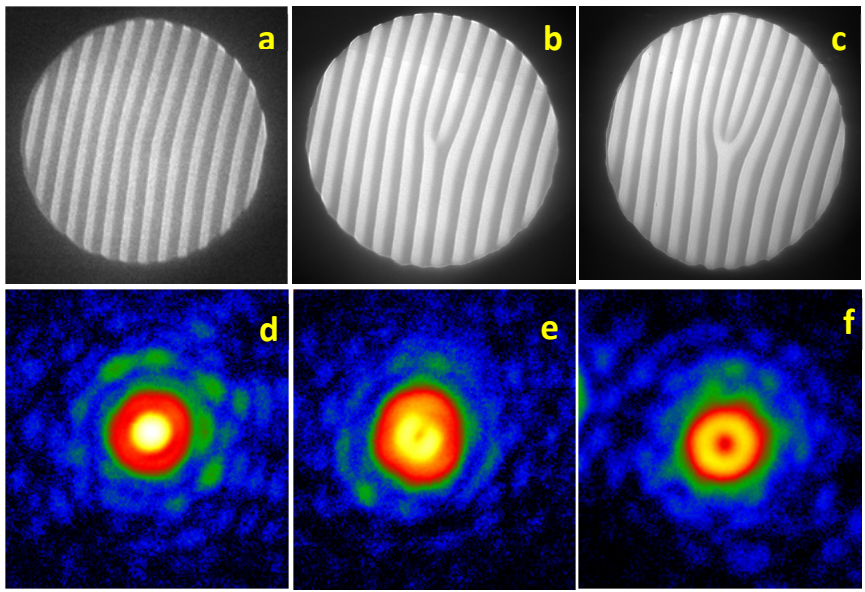

**Figure 5.** Beam generation by small-aperture phase holograms. **a**, **b**, and **c** show in-focus bright-field images of phase holograms with small aperture radii $R = 1.22\,\mu\text{m}$, convergence angles $\alpha = 1\,\mu\text{rad}$ and respective topological charges $n = 0$, 1 and 2. **d**, **e** and **f** show the experimental intensity patterns obtained from these respective holograms. Notably, the hologram with $n = 0$ gives rise to a single, well-defined point of maximum beam intensity, whereas higher topological charges lead to doughnut-shaped intensity patterns.

## Propagation

In order to characterize the effective propagation range of the Bessel beams generated using our technique, we examined the intensity at beam center for the first diffracted order, in the case $n = 0$, i.e. for an electron beam carrying zero OAM. The holograms used in this experiment featured large aperture radii, and resembled the holographic mask shown in Figure 3-a. The intensity values thus obtained are shown as a function of propagation distance in Figure 6, along with theoretical plots obtained from simulations.

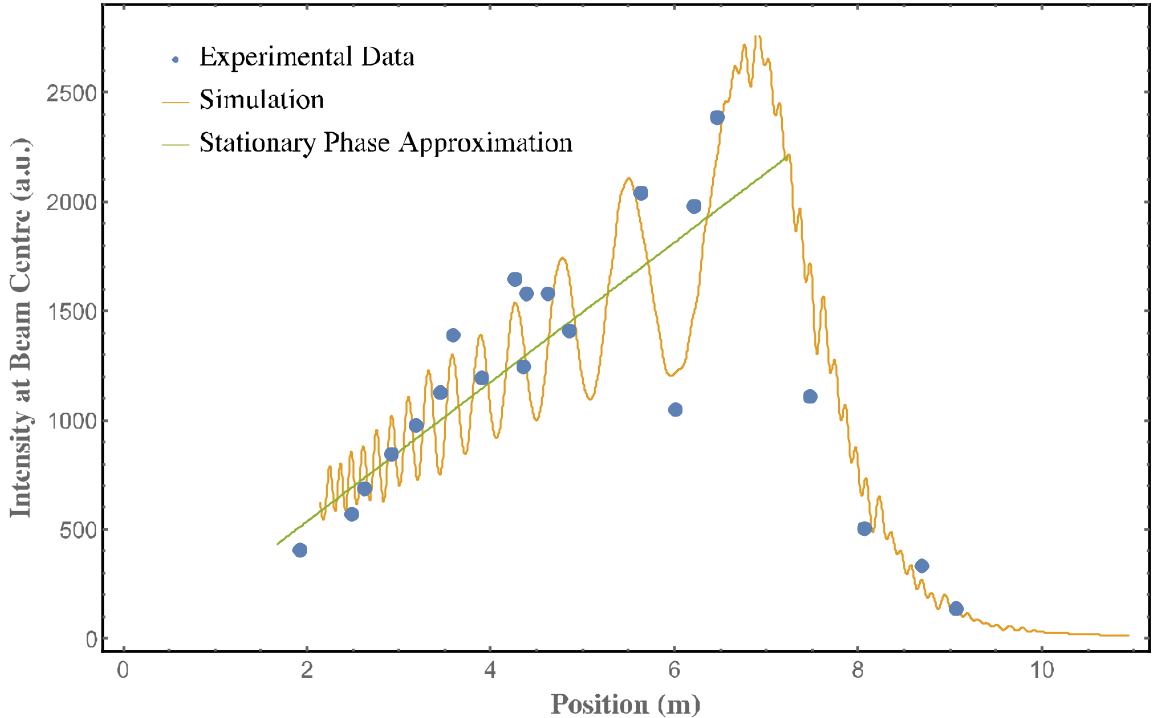

**Figure 6.** Intensity at beam center as a function of propagation distance. Simulated (black curve) and experimental (red points) intensities on beam center as a function of propagation for an electron Bessel probe with parameters $n = 0$, $\frac{k_\rho}{k} = 6 \ \mu rad$, and $R = 5 \ \mu m$.

Our results indicate that, apart from some oscillations, the beam intensity rises to a maximum value at $z_{max} = 0.7$ m. Although the stationary phase approximation (SPA) fails to predict the intensity oscillations observed from $z = 0.2$ m to $z_{max}$, the initial, linear increase in average intensity with $z$, and its rapid disappearance after $z_{max}$, correspond closely to the behaviour prescribed by the SPA.

We note also that it can be difficult to identify the plane at which the Fraunhofer condition is satisfied when carrying out experiments involving small aperture radii $R$. By definition, the Fraunhofer plane is the position at which the $0^{th}$ diffraction order of the transmitted beam is most tightly focused. However, when $R$ is small, it is in practice difficult to clearly identify the zeroth-diffracted order in beam cross-section images obtained experimentally. Further complicating matters, each diffraction order is focused at a different position, so that an unambiguous identification of the Fraunhofer plane is challenging to achieve. Notwithstanding these limitations, techniques have been developed that allow the zeroth diffracted order to be identified, by deliberately introducing a condenser astigmatism to the beam, as reported in reference [29].

## Phase Description

Since vortex beams are most completely described by referring to their transverse phase structure, a great deal of emphasis has been placed on the development of techniques that might allow for the retrieval of phase information from such beams [30]. For our purposes, a realistic reconstruction of the phase of the electron beam can be achieved from calculations based on experimentally measured hologram profiles. Since the wavefunction of a transmitted electron beam can be determined from the hologram thickness profile, the beam structure can be calculated at any propagation plane, using the techniques discussed earlier.

In a previous study [6], we demonstrated that when beam coherence effects are accounted for, a very good agreement exists between the modeled electron wavefunction, and the beam's experimentally observed intensity distribution. Thus, this technique provides an initial, indirect means by which to characterize the transmitted electron beams. Transverse intensity and phase profiles calculated for an electron beam carrying an OAM of $n = 1$, generated by a small aperture, are shown in Figure 7. Given that the intensity pattern calculated for the beam corresponds closely to those obtained experimentally, we assume that the calculated phase distribution represents an accurate picture of the beam phase structure as well.

We also carried out a simple phase analysis, analogous to that reported in [31], to locate beam phase vortices. Our results show that, in the case $n = 2$, in figure 5c the second-order vortex decomposes into two separate vortices of first order, as predicted in [32]. This observation cannot be ascribed to lensing effects, owing to the fact that this phenomenon is not accounted for by our simulation technique. Rather, we believe this decomposition to arise from imperfections in the grating [33,34]

If only the OAM content of the first-order beam is of interest, a more direct characterization of the first diffracted order can be achieved by interfering the first-order diffracted beam with the zeroth-order as a reference. The resulting pitchfork-shaped interference pattern produces a vortex dislocation that indicates the OAM content of the first-order beam.

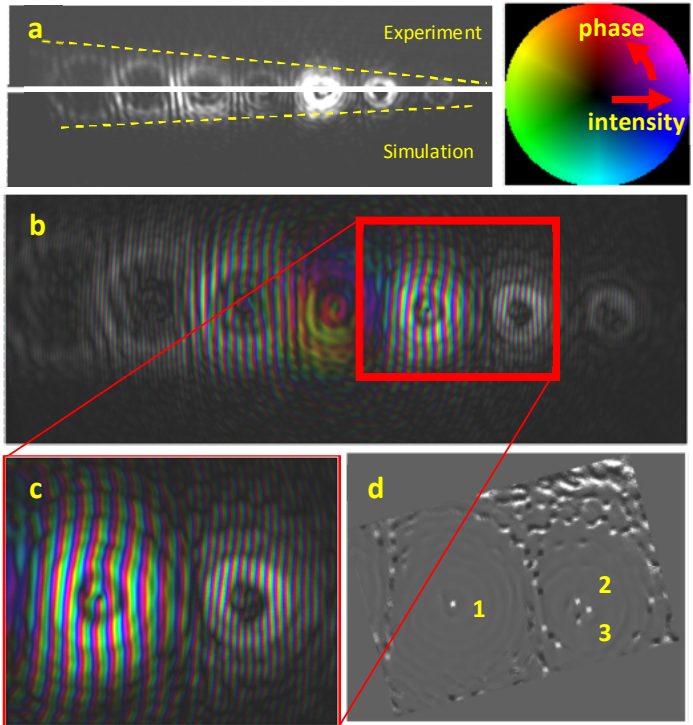

**Figure 7.** Correspondence between experimental and theoretically anticipated beam phase structures. **a** Simulated and experimental beam propagation, showing agreement at $z = 0.1$m. From this propagation distance and the known hologram profile, it is possible to reconstruct the phase structure of the beam. **b** Orders of diffraction obtained from a phase hologram with parameters $n = 1$, $\frac{k_\rho}{k} = 1\ \mu rad$, and $R = 1.22\ \mu rad$, with superimposed phase structures. In the figure, beam phase is indicated by hue, and intensity by brightness. **c** Enlarged view of the first and second diffracted orders shown in part **b**. **d** Reconstruction of the positions of phase vortices in the original beam (see Appendix III).

In Figure 8, we show beam cross-sectional images obtained for several diffracted orders at various effective propagation distances about the $m = 0$ order focal point. As can be gathered from the figure, every diffracted order is found to focus at a different location. Further, the sizes of the diffracted beams are found to depend linearly on the indices $m$ of the respective diffracted orders (a consequence of the conical shape of the beam), in agreement with the anticipated range of validity of Equation (9) (See Appendix I), $\rho < R - mk_\rho z/k$. It can be readily be seen that the angle $\beta$ is related to $\alpha$ through the angular separation of the order, $\theta_B$ ( proportional to δ in fig 8) , so that $\beta \theta_B = \alpha$.

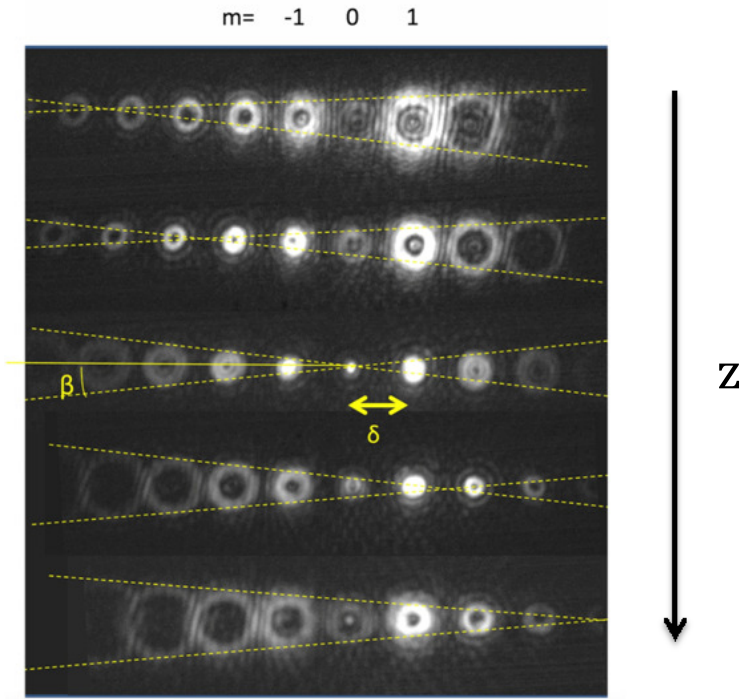

**Figure 8.** Propagation and focal characteristics of various diffracted orders. Experimentally obtained intensity profiles associated with various orders of diffraction, which are visibly focused at different propagation distances. The anticipated linear dependence of beam size on diffraction order is verified by equidistantly pasting together transverse intensity profiles for each order. We note that, as such, the figure does not show any diffractive spreading of the various orders upon propagation.

**Bessel Probes in STEM**

So far, we have considered the characteristics of Bessel beams as probed in the Lorentz mode. However, some of the most interesting properties of Bessel beams can be exploited only if they are used to illuminate a sample in STEM mode.

If the hologram is placed at the condenser aperture, the condenser lens (and pre-field of the objective) produces a demagnification of the condenser aperture, and an angular magnification of the beam. As a result, the hologram can easily reach nanometer and sub-nanometer sizes.

An image of the hologram is shown in Figure 9-a, for a probe size of $0.5$ nm. The Fresnel diffraction from the hologram is shown in Figure 9-b. In this case, the rings corresponding to most diffraction orders are shown to be partially overlapped. The beam convergence on the sample plane is $1.9$ mrad in this case. The microscope, thus employed, has produced an angular magnification of 3 orders of magnitude.

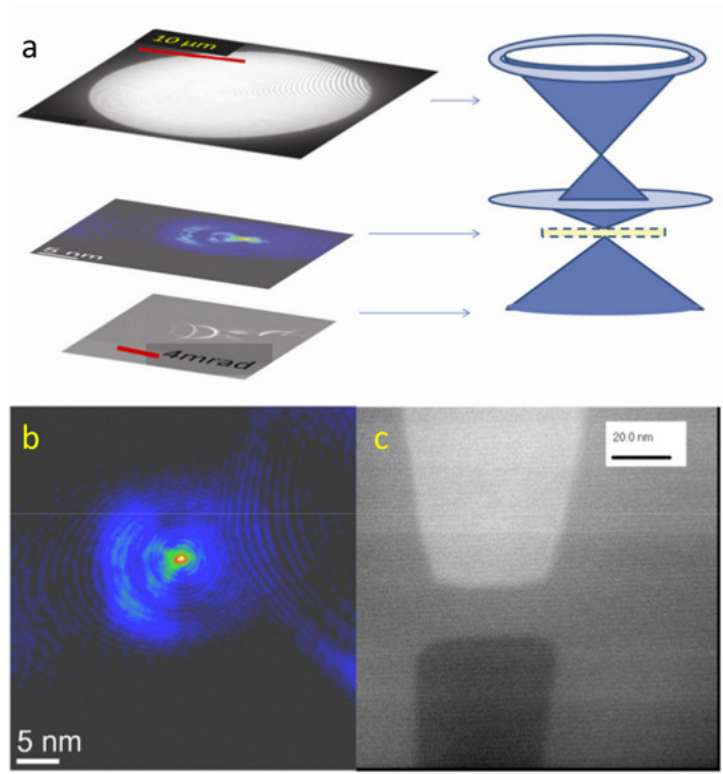

**Figure 9.** Imaging with STEM in Fresnel mode. **a** Diagram of the Fresnel operating mode showing the phase hologram in the condenser aperture, the near-field Bessel beam intensity profile in the sample plane, and far-field ring-like profile for a beam generated from the hologram shown. The phase hologram was prepared with parameters $n = 0$, $\frac{k_\rho}{k} = 3\ \mu rad$, and $R = 10\ \mu m$. **b** Image of the transverse intensity profile of an electron Bessel beam **c** STEM image generated using the electron beam shown in part **a**.

Using the Bessel beams in Figure 8-a, we obtained the STEM image shown in Figure 8-c, which demonstrates that a sample (in this case, a Si–SiO STI structure) can be scanned with sub-2 nm resolution, in spite of the presence of diffraction orders for which $m \neq 1$. This is made possible because increases in $k_\rho$ result in a greater delocalization of diffraction orders for which $m \neq 1$ at the $m = 1$ focal plane, meaning that only the first-order beam is properly focused. The influence of cross-contamination by other diffracted orders is therefore negligible, and imaging can be carried out as if working with a beam entirely consisting of the pure first diffracted order. As expected, the distance to the focal point increases with the value of $k_\rho$, so that this parameter must be considered in optimizing STEM experiments.

**Future Applications of Bessel Beams in STEM**

Our experiment has exhibited a maximum magnification of up to 1000X. One might imagine that the degree of magnification achieved by the electron Bessel beams described herein could be improved simply by increasing $k_\rho$, due to the corresponding decrease in beam central spot size prescribed by Equation (9). However, increases in $k_\rho$ are met by corresponding increases in the focal point of the first diffracted order. In order to ensure that the beam remains focused on the sample, it then becomes necessary to compensate for this effect by reducing lens excitation, therefore decreasing the demagnification factor of the system. The compensatory measures designed to re-focus the electron beam therefore ultimately cancel any benefit that might otherwise arise from an increase in the transverse wavenumber $k_\rho$. This limitation could certainly be overcome by a more flexible illumination scheme, however. We note also that increasing the size of the hologram aperture would not improve the quality of images obtained in our configuration. As discussed previously, the beam convergence is not determined by the size of the aperture in the Fresnel regime.

In order to emphasize the possible advantages of Bessel probes, one can calculate the STEM annular dark field (ADF) transfer function for the electron beam in the absence of aberration. The STEM ADF transfer function $H(k)$ is given by the Fourier transform of the probe intensity [35]:

$$H(k) = A(k) \otimes A^*(k). \quad (12)$$

Here, we define $A(k)$ as the far-field amplitude of the beam, which in our case is ideally given by an infinitely narrow ring of diameter $k_\rho$, as per Equation (10), assuming a perfect Bessel beam, and as before denote by $\otimes$ the convolution operator. In this case, the above convolution itself yields a delta function centered at $k_\rho$. In practice, however, $A(k)$ will take nonzero values within a range $\Delta k/2$ of its maximum value at $k_\rho$, so that the transfer function $H(k)$ will itself exhibit some finite spread about its maximum value.

Figure 10 shows the transfer function obtained for a microscope using a conventional probe with 15 mrad convergence (with and without aberration), and a Bessel probe. The transfer functions have been scaled so as to have the same values at the in the lowest frequency region. Notably, the Bessel probe confers an advantage over other probes in the high frequency region, just above $\sim 1 A^{-1}$. In lower spatial frequency regions, however, the Bessel probe is clearly non-ideal.

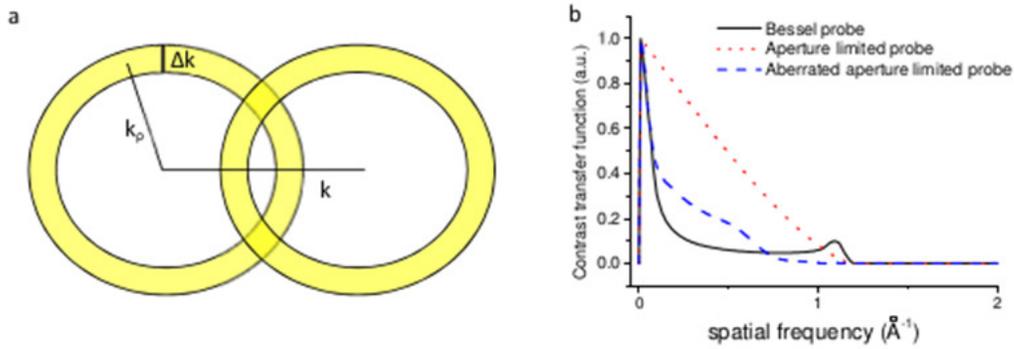

**Figure 10.** Bessel beam transfer functions. Comparison of aperture-limited (with and without aberrations) and Bessel probe contrast transfer functions. The plot has been normalized so as to ensure that all probes display equal transfer functions at the lowest spatial frequency. Although the aperture limited probe is best suited to imaging at most spatial frequencies, the Bessel probe is found to exhibit superior performance above a frequency of $1A^{-1}$. The results displayed were obtained by simulating beams for which $n = 0$, and actual convergence after the lens of 15 mrad. For the Bessel probe, $\frac{\Delta k}{k} = 0.06$ (solid line). For the aperture limited probe, we considered the probe with and without aberration (dotted line). For the aberrated case, we considered a spherical aberration coefficient Cs=0.5 mm, and a defocus of 40 nm (dashed curve).

The high value of the contrast transfer function associated with the Bessel probe immediately above the $1A^{-1}$ spatial frequency may in part be attributed to the small central spot size of the Bessel probe. More significant, however, is the presence of the evenly spaced Bessel probe rings, which account for the majority of this super-resolution effect. The rings are spaced at intervals on the order of the central spot size of the Bessel beam. When the ring spacing matches the spatial frequency of a feature of interest, each ring contributes materially to the resolution achieved, so that an image can be obtained.

In order to demonstrate that the superresolution achieved using Bessel beams in the indicated spatial frequency domain does not primarily arise from the small size of the central maximum of the beam, we can consider a beam with $n > 0$, for which no beam intensity is found at beam center. For small $n$ the function $A(k)$ will approximately take the form of a ring, meaning that the transfer function $H(k)$ will not qualitatively differ from that obtained for the $l = 0$ case, in which the majority of beam intensity is on or near the optical axis.

We note in closing that Bessel beams also benefit from insensitivity to chromatic aberration. Chromatic aberration arises from differences in the focal positions associated with different de Broglie wavelengths. Since Bessel beams are propagation invariant in the near-field regime, and therefore do not focus for any wavelength, the differences in foci that lead to chromatic aberration do not play a role for these non-diffracting beams.

**Conclusion**

We have explored the theory of Bessel beam holographic generation in detail, examining the impact and importance of hologram parameters such as the groove shape and depth, aperture size, fringe spacing and modulation. By optimizing these parameters, we have experimentally achieved Bessel beam generation with efficiencies as high as 37±3%. Moreover, we have demonstrated experimentally the successful generation of Bessel beams characterized by variable transverse wavenumbers, topological charges and ranges of non-diffractive propagation through direct measurement and observation of beam structure. We have shown that these beams can be used to image samples with an effective convergence of 1.9 mrad. This remarkable result is made possible due to the differing focal characteristics of each diffracted order, which allow only one order to be focused at any given position at a time. Finally, as a preliminary demonstration of the versatility and potential of the electron beams generated using our technique, we manufactured and mounted a hologram in the condenser aperture of an STEM, scanning a sample with nanometer resolution. We believe that this systematic study will greatly facilitate the application of Bessel beams to imaging and electron microscopy.

## APPENDIX I: Fresnel Propagation of Diffracted Electron Beams

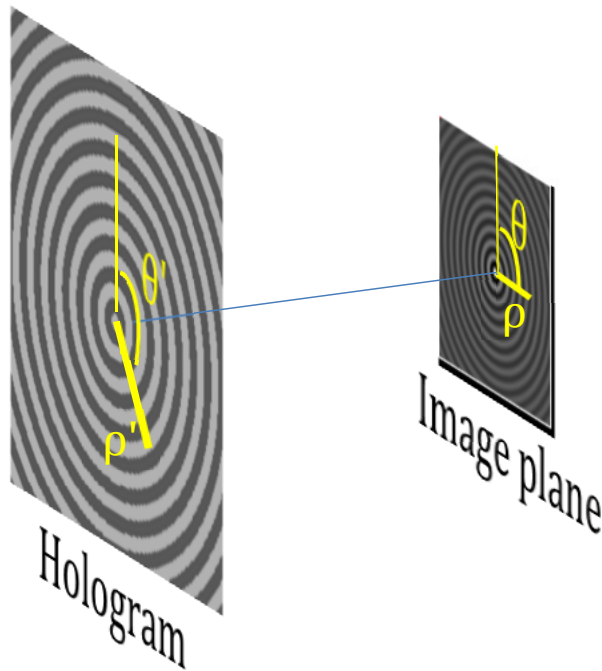

**Figure 11.** Schematic representation of a Bessel phase hologram and beam image plane for the purpose of determining near-field and far-field beam profiles. Cylindrical coordinates are used to indicate points in both planes. Primed coordinates refer to the hologram plane, while unprimed coordinates refer to the image plane.

The wavefunctions associated with each diffraction order will evolve through free space propagation, beyond the holographic plate. The effect of this propagation can be calculated by evaluating the Fresnel diffraction integral

$$\Psi_t^m(\rho,\phi,z) = \frac{e^{ikz}}{i\lambda z}\iint dx'dy'\, A_m(x',y')e^{-i\frac{k}{2z}((x-x')^2+(y-y')^2)}, \quad (A1)$$

where $A_m(x',y')$ is the aperture function, which describes the phase and amplitude modulation induced in an incident beam by the holographic phase mask. With respect to the $m^{th}$ diffraction order, the aperture function will take the form $A_m(x,y) \to A_m(\rho,\phi) = e^{im(k_\rho \rho + n\phi)}$. Hence, the diffraction integral becomes

$$\Psi_t^m(\rho,\phi,z) = \frac{e^{ik(z-\rho^2/2z)}}{i\lambda z}\int_0^R d\rho'\rho' e^{i(mk_\rho \rho' - k(\rho')^2/2z)} \int_0^{2\pi} d\phi'\, e^{imn\phi'} e^{ik(\rho\rho'/z)\cos(\phi-\phi')},$$

where the limits of the outer integral reflect the finite aperture of the holographic mask. Applying the Jacobi-Anger expansion to the integral over $\phi$ yields

$$\int_0^{2\pi} d\phi'\, e^{imn\phi'} e^{ik(\rho\rho'/z)\cos(\phi-\phi')} = \int_0^{2\pi} d\phi'\, e^{imn\phi'} \sum_{p=-\infty}^{\infty} i^p J_p(k(\rho\rho'/z)) e^{ip(\phi-\phi')}$$

And with $\int_0^{2\pi} d\phi'\, e^{i(mn-p)\phi'} = 2\pi \delta_{mn,p}$, this gives

$$\int_0^{2\pi} d\phi'\, e^{imn\phi'} e^{ik(\rho\rho'/z)\cos(\phi-\phi')} = 2\pi i^{mn} e^{imn\phi} J_{mn}(k(\rho\rho'/z))$$

So that we have

$$\Psi_t^m(\rho,\phi,z) = -\frac{2\pi i^{mn+1}}{\lambda z} e^{ik\left(z-\frac{\rho^2}{2z}\right)} e^{imn\phi} \int_0^R d\rho'\rho' e^{i\left(mk_\rho \rho' - \frac{k(\rho')^2}{2z}\right)} J_{mn}\left(k\left(\frac{\rho\rho'}{z}\right)\right) \quad (A2)$$

We now turn to address a related problem, namely that of evaluating the integral

$$\int_0^R F(x)e^{ig(x)}dx, \quad (A3)$$

where $g(x)$ is taken to represent a function with a single extremum, at $x = x_c$, such that $\frac{d}{dx}g(x_c) = 0$. We note that the exponential term $e^{ig(x)}$ oscillates rapidly with $x$, except at $x_c$, where the derivative of $g(x)$ vanishes. As a result, only the extremum $x_c$ of $g(x)$ will contribute materially to the integral (A3). The integral may therefore be approximated by taking $g(x) \approx g(x_c) + \frac{1}{2!}\frac{d^2 g(x)}{dx^2}\big|_{x=x_c}(x-x_c)^2$, and $F(x) = F(x_c)$ so that

$$\int_0^R F(x)e^{ig(x)}dx \approx e^{ig(x_c)}F(x_c)\int_0^R e^{\frac{i}{2!}\frac{d^2 g(x)}{dx^2}\big|_{x=x_c}(x-x_c)^2}dx$$
$$= e^{ig(x_c)}F(x_c)\int_{x_c}^{R+x_c} e^{\frac{i}{2!}\frac{d^2 g(x)}{dx^2}\big|_{x=x_c}x^2}dx$$
$$= e^{ig(x_c)}F(x_c)\int_{x_c}^{R+x_c} e^{i\alpha x^2}dx, \quad (A4)$$

where $\alpha \equiv \frac{1}{2}\frac{d^2 g(x)}{dx^2}\big|_{x=x_c}$. We note that the integral (A4) is mathematically equivalent to the expression (A2) that we wish to evaluate, if we choose $F(x) = xJ_{mn}(\gamma x)$, with $\gamma \equiv \frac{k\rho}{z}$, and $g(x) = mk_\rho x - \frac{kx^2}{2z}$. In this case, we find that $x_c = mk_\rho z/k$ and $\alpha(z) = -\frac{k}{2z}$, whence

$$\int_0^R d\rho' \rho' e^{i\left(mk_\rho \rho' - \frac{k(\rho')^2}{2z}\right)}J_{mn}\left(k\left(\frac{\rho \rho'}{z}\right)\right) \approx \frac{mk_\rho z}{k}J_{mn}(mk_\rho \rho)e^{i\left(\frac{m^2 k_\rho^2 z}{2k}\right)}I(R,z), \quad (A5)$$

where we have defined $I(R,z) \equiv \int_{x_c}^{R+x_c} e^{i\alpha(z)x^2}dx$, $R$ representing the radius of the holographic mask. The approximate equality (A5) can now be substituted into (A2) to yield

$$\Psi_t^m(\rho,\phi,z) \approx -i^{mn+1}mk_\rho e^{i\left(kz - \frac{k\rho^2}{2z} + mn\phi + \frac{m^2 k_\rho^2 z}{2k}\right)}J_{mn}(mk_\rho \rho)I(R,z). \quad (A5)$$

The expression for $\Psi_t^m(\rho,\phi,z)$ provided above includes a contribution from the aperture limiting function $I(R,z)$, which accounts for the finite size of the phase hologram. As $R \to \infty$, this integral effectively becomes Gaussian, whereas for all other aperture radii, $I(R,z)$ takes the form of an error function. In either case, the term $I(R,z)$ does not affect the transverse profile $\psi_t^m(\rho,\phi)$ of the beam, so that

$$\psi_t^m(\rho,\phi) = Ne^{i\left(kz - \frac{k\rho^2}{2z} + mn\phi + \frac{m^2 k_\rho^2 z}{2k}\right)}J_{mn}(mk_\rho \rho), \quad (A6)$$

where $N$ is a normalization constant. We note also that (A5) is contingent upon the approximation (A4), and therefore holds true only to the extent that $\frac{d}{d\rho}g(\rho) \approx 0$, or equivalently, when

$$z \approx \frac{k\rho}{mk_\rho}. \quad (A7)$$

Hence, the transmitted beam associated with the $m^{th}$ diffraction order can be considered to represent a Bessel beam of $m^{th}$ order only within the range of $\rho$ and $z$ prescribed by (A7). This condition can be satisfied for values of $z$ such that $z \leq \frac{kR}{mk_\rho}$. For other values of $z$, the approximation (A4) fails, so that the Bessel character of the transmitted beam is no longer maintained. We note also that the range of $z$ over which equation (A6) holds true is also limited by the range of validity of the Fresnel approximation (A1). Hence, equation (A6) can adequately represent the transmitted wavefunction only in the range $\rho \ll z \leq \frac{kR}{mk_\rho}$.

We note also that the oscillatory behaviour of the Bessel function component in the integrand of equation (A2) can contribute significantly when the Bessel function's argument becomes large. Indeed, for large $x$,

$$J_{nm}(x) \sim \sqrt{\frac{2}{\pi x}} \cos\left(\frac{nm\pi}{2} + \frac{\pi}{4} - x\right).$$

Therefore, accounting for the oscillatory behaviour of the integrand (A2) results in the inclusion of an additional phase term $k\left(\frac{\rho \rho'}{z}\right) - \frac{nm\pi}{2} - \frac{\pi}{4}$ in the expression for $g(\rho')$, so that $\frac{d}{d\rho'}g(\rho') = mk_\rho - \frac{k\rho'}{z} + k\left(\frac{\rho}{z}\right)$. Consequently, the condition for stationary phase becomes $\rho' = \rho + mk_\rho z/k$. However, since the integral (A2) has an upper bound at $\rho' = R$, the stationary phase condition can be met only for $\rho < R - mk_\rho z/k$, so that the generated beam will take Bessel form only in this range.

### APPENDIX II: Far-Field Propagation

We determine the wavefunction of the $m^{th}$ diffraction order in the far-field by evaluating the Fraunhofer diffraction integral

$$\psi_t^m(k_x, k_y) = N \iint dx'dy' A_m(x', y') e^{i(k_x x' + k_y y')}. \quad (A8)$$

Following a procedure analogous to that outlined in Appendix I, we choose $A_m(x, y) \to A_m(\rho, \phi) = e^{im(k_\rho \rho + n\phi)}$, so that

$$\psi_t^m(k_x, k_y) = N \int_0^R d\rho' \rho' e^{imk_\rho \rho'} \int_0^{2\pi} d\phi' e^{imn\phi'} e^{i\rho'(k_x \cos\phi' + k_y \sin\phi')}.$$

But since $k_x = k \cos\phi$ and $k_y = k \sin\phi$, this becomes

$$\psi_t^m(k_\rho) = N \int_0^R d\rho' \rho' e^{imk_\rho \rho'} \int_0^{2\pi} d\phi' e^{imn\phi'} e^{i\rho' k \cos(\phi-\phi')}.$$

We address the inner integral first, once again resorting to the Jacobi-Anger identity to write

$$\int_0^{2\pi} d\phi' e^{imn\phi'} e^{i\rho' k \cos(\phi-\phi')} = \int_0^{2\pi} d\phi' e^{imn\phi'} \sum_{p=-\infty}^{\infty} i^p J_p(k\rho') e^{ip(\phi-\phi')},$$

so that, with $\int_0^{2\pi} d\phi' e^{i(mn-p)\phi'} = 2\pi \delta_{mn,p}$,

$$\int_0^{2\pi} d\phi' e^{imn\phi'} e^{i\rho' k \cos(\phi-\phi')} = 2\pi i^{mn} e^{imn\phi} J_{mn}(k\rho').$$

As a result, the transverse wavefunction is then given by

$$\psi_t^m(k_\rho) = \tilde{N} e^{imn\phi} \int_0^R d\rho' \rho' e^{imk_\rho \rho'} J_{mn}(k\rho').$$

We now note that the above expression may be written equivalently in the form

$$\psi_t^m(k_\rho) = -i\tilde{N} e^{imn\phi} \frac{\partial}{\partial(mk_\rho)} \int_0^R d\rho' \, e^{imk_\rho \rho'} J_{mn}(k\rho'). \quad (A9)$$

The integral on the right hand side of this new equation can be further decomposed by making use of the Euler identity:

$$\int_0^R d\rho' e^{imk_\rho \rho'} J_{mn}(k\rho')$$
$$= \int_0^R d\rho' \cos(mk_\rho \rho') J_{mn}(k\rho') + i \int_0^R d\rho' \sin(mk_\rho \rho') J_{mn}(k\rho').$$

The above expression takes on qualitatively different solutions, depending on the relative values of $k_\rho$ and $k$. For $k_\rho < k$, and taking $R \to \infty$ as the upper bound of the integral (which is reasonable for $kR \gg 1$), we have

$$\int_0^\infty d\rho' e^{imk_\rho \rho'} J_{mn}(k\rho') = \frac{1}{\sqrt{(k^2 - m^2 k_\rho^2)}} e^{mn \sin^{-1}\left(\frac{mk_\rho}{k}\right)},$$

whereas for $k_\rho > k$,

$$\int_0^\infty d\rho' e^{imk_\rho\rho'} J_{mn}(k\rho') = i\frac{k^{mn}}{\sqrt{m^2k_\rho^2 - k^2}}\left(mk_\rho + \sqrt{m^2k_\rho^2 - k^2}\right)^{-mn} e^{i\frac{mn\pi}{2}}.$$

Upon substitution of this expression into (A9), and evaluation of the corresponding derivative, we finally obtain the transverse wavefunctions

$$\psi_t^m(k_\rho) = \tilde{N}e^{imn\phi}\frac{1}{(k^2 - m^2k_\rho^2)^2}\left(-ik_\rho\sqrt{k^2 - m^2k_\rho^2} + m(k^2 - m^2k_\rho^2)\right)e^{im\sin^{-1}(mk_\rho/k)}$$

for $k_\rho < k$, and

$$\psi_t^m(k_\rho) = \tilde{N}e^{imn\phi}\frac{k^m}{(k^2 - m^2k_\rho^2)^2}\left(mk_\rho + \sqrt{k^2 - m^2k_\rho^2}\right)^{-m}\left(m(k^2 - m^2k_\rho^2) - mk_\rho\sqrt{m^2k_\rho^2 - k^2}\right)e^{i\frac{m\pi}{2}}$$

for $k_\rho > k$. In either case, we have that $\psi_t^m(\rho, \phi) \to \infty$ for $mk_\rho \to k$, so that, upon normalization, the far-field wavefunction becomes a delta function, centered at $k_\rho = k$, i.e. $\psi_t^m(k_\rho) = e^{imn\phi}\delta(k - mk_\rho)$.

**APPENDIX III: Locating Phase Singularities**

Phase singularities occur for non-vanishing integer winding numbers $n$, as defined by closed contour integrals of the form

$$2\pi n = \oint \vec{\nabla}\varphi \cdot d\vec{s}, \quad (A10)$$

where $\varphi$ is the phase of the wavefunction, and $d\vec{s}$ is an infinitesimal segment of the closed path. Here, $\vec{\nabla}\varphi$ is obtained from the periodic derivative, defined for example in [36], as

$$\frac{\delta\varphi}{\delta x} = -ie^{-i\varphi}\frac{\partial}{\partial x}(e^{-i\varphi}).$$

This definition removes the artificial $2\pi$ discontinuity that has no impact on the phase.

In practice, the integral (A10) was evaluated over small rectangular paths, typically of dimension 5x5 pixels. The results were then entered into a map of the beam, and the integral was found to yield zero within floating point precision, and $2\pi n$ about the vortices [37].